\documentclass[conference]{IEEEtran}
\usepackage{cite}
\usepackage{amsmath,amssymb,amsfonts}
\usepackage{algorithmic}
\usepackage{graphicx}
\usepackage{textcomp}
\usepackage{xcolor}
\usepackage{multirow}

\ifCLASSINFOpdf
\else
\fi
\usepackage{url}

\begin{document}
%
\title{Automated Detection of GDPR Disclosure Requirements in Privacy Policies using Deep Active Learning}

\author{\IEEEauthorblockN{Tamjid Al Rahat}
\IEEEauthorblockA{University of Virginia\\
tr9wr@virginia.edu}
\and
\IEEEauthorblockN{Tu Le}
\IEEEauthorblockA{University of Virginia\\
tnl6wk@virginia.edu}
\and
\IEEEauthorblockN{Yuan Tian}
\IEEEauthorblockA{University of Virginia\\
yt2e@virginia.edu}}

\maketitle

\begin{abstract}
Since GDPR came into force in May 2018, companies have worked on their data practices to comply with this privacy law. In particular, since the privacy policy is the essential communication channel for users to understand and control their privacy, many companies updated their privacy policies after GDPR was enforced. However, most privacy policies are verbose, full of jargon, and vaguely describe companies' data practices and users' rights. Therefore, it is unclear if they comply with GDPR. In this paper, we create a privacy policy dataset of 1,080 websites labeled with the 18 GDPR requirements and develop a Convolutional Neural Network (CNN) based model which can classify the privacy policies with an accuracy of 89.2\%. We apply our model to perform a measurement on the compliance in the privacy policies. Our results show that even after GDPR went into effect, 97\% of websites still fail to comply with at least one requirement of GDPR.
\end{abstract}


%

\section{Introduction} \label{sec:intro}
Privacy regulations are introduced to protect the personal data of individuals that are collected by public or private companies, governments, and other individuals. Among many privacy regulations, GDPR (\textit{General Data Protection Regulation}) has been considered as one of the strictest privacy regulations~\cite{gdpr-strict1,gdpr-strict2}. The primary purpose of GDPR is to give more control to individuals over their personal data and to ensure their rights regarding those data. GDPR applies to any individual or company that collects, stores, or processes any personal information in connection to services or goods offered in European Union (EU) countries. It significantly affects millions of websites, applications, and online services. 

Since GDPR came into effect in May 2018, companies have been focusing on reforming their data practices, including changing the privacy policies~\cite{policy-changes}, in view of the fact that non-compliance with GDPR could be fined up to 4\% of the company's total annual revenue or 20 million Euros, whichever is higher (article 83~\cite{gdpr}). However, even with such stiff fines and penalties, it appears that many companies are not yet fully compliant with GDPR. For example, Information Commissioner's Office (ICO) of the United Kingdom 
fined British Airways 183 million Euros~\cite{ico-fine} for failing to protect users' personal data and violating GDPR. The Irish Data Protection Commission (DPC) has opened 19 investigations~\cite{irish-dpc} into big companies' potential privacy breaches (e.g., Google~\cite{irish-dpc-google}, Facebook~\cite{irish-dpc-facebook}, and Twitter~\cite{irish-dpc-twitter}) since May 22, 2019. 

One of the most important GDPR requirements is that companies must disclose to users their personal data handling process and certain rights of the individuals (e.g., disclose information about what personal data they collect, how they collect, store, process, and share it). 
Since the privacy policy is the de facto standard mechanism for communicating companies' data practices~\cite{reidenberg2015disagreeable
}, studying the privacy policies provides useful insights to understand the GDPR compliance. 
In this paper, we explore the research question: \emph{Do privacy policies of the companies fully comply with GDPR requirements to communicate their data practices and users' rights?} To answer the question, we build a Convolutional Neural Network (CNN) based classifier to determine whether privacy policies are fully compliant with GDPR. One of the major challenges for training such a classifier is to get an annotated dataset. Unfortunately, as far as we know, there is no publicly available privacy policy dataset labeled with the GDPR requirements. Therefore, we create a privacy corpus of 9,761 privacy policies from top websites by using UK-based VPNs. To understand what is required to be disclosed in the privacy policies by GDPR, we identify 18 requirements (Table ~\ref{tab:category-statistics}) from the GDPR legislation~\cite{gdpr_legislation} with the help of two legal experts, who also train four human annotators to annotate privacy policies. These 18 categories of information must be disclosed in the privacy policy when companies (i.e., data controllers) collect personal data from human subjects or other sources. We randomly select 1,080 privacy policies from our corpus to annotate using 18 labels that represent the requirements of GDPR. To the best of our knowledge, we are the first to contribute a privacy policy dataset labeled with GDPR requirements. Another challenge is that privacy policy segments of different GDPR requirements may contain overlapping features that can lead to a lot of misclassification. To overcome this challenge, we use CNN, which can detect the set of features that contribute most towards a class regardless of their position in the input. Our initial trained model achieves an accuracy of 80.5\%. To further improve accuracy, we need more labeled data. However, collecting labeled data is expensive and time-consuming, whereas unlabeled data are free. Therefore, we leverage pool-based \textit{active learning} technique to improve our model's accuracy using less labeled data. Our final classification model achieves an accuracy of 89.2\%, which is a 10.8\% relative increase compared to the initial performance.

We apply our classification model to determine the compliance of 9,761 top websites' privacy policies with the 18 GDPR requirements. We find that there are only 3\% of companies fully comply with GDPR in their privacy policies. We also identify six major requirement categories of GDPR  implemented only by 15-20\% websites. For example, GDPR requires companies to disclose any information regarding profiling or any other automated decision making by using users' personal data. Surprisingly, only 15.3\% of companies disclose information about this in their privacy policies.

In this paper, we make the following contributions:
\begin{itemize}
    \item We contribute a new dataset containing 1080 privacy policies annotated by skilled annotators using labels representing 18 privacy disclosure requirements of GDPR. We will make our annotated dataset publicly available.
    \item We develop a CNN-based privacy policy classifier to classify privacy policy segments into 18 GDPR requirements. Our model achieves an accuracy of 89.2\% with an average F1-score of 0.88. With the aid of our model, we investigate 9,761 online privacy policies, which shows that only 3\% fully comply with GDPR.
\end{itemize}


\section{Overview} \label{sec:overview}
\begin{figure}[!t]
\centering
\includegraphics[width=\linewidth]{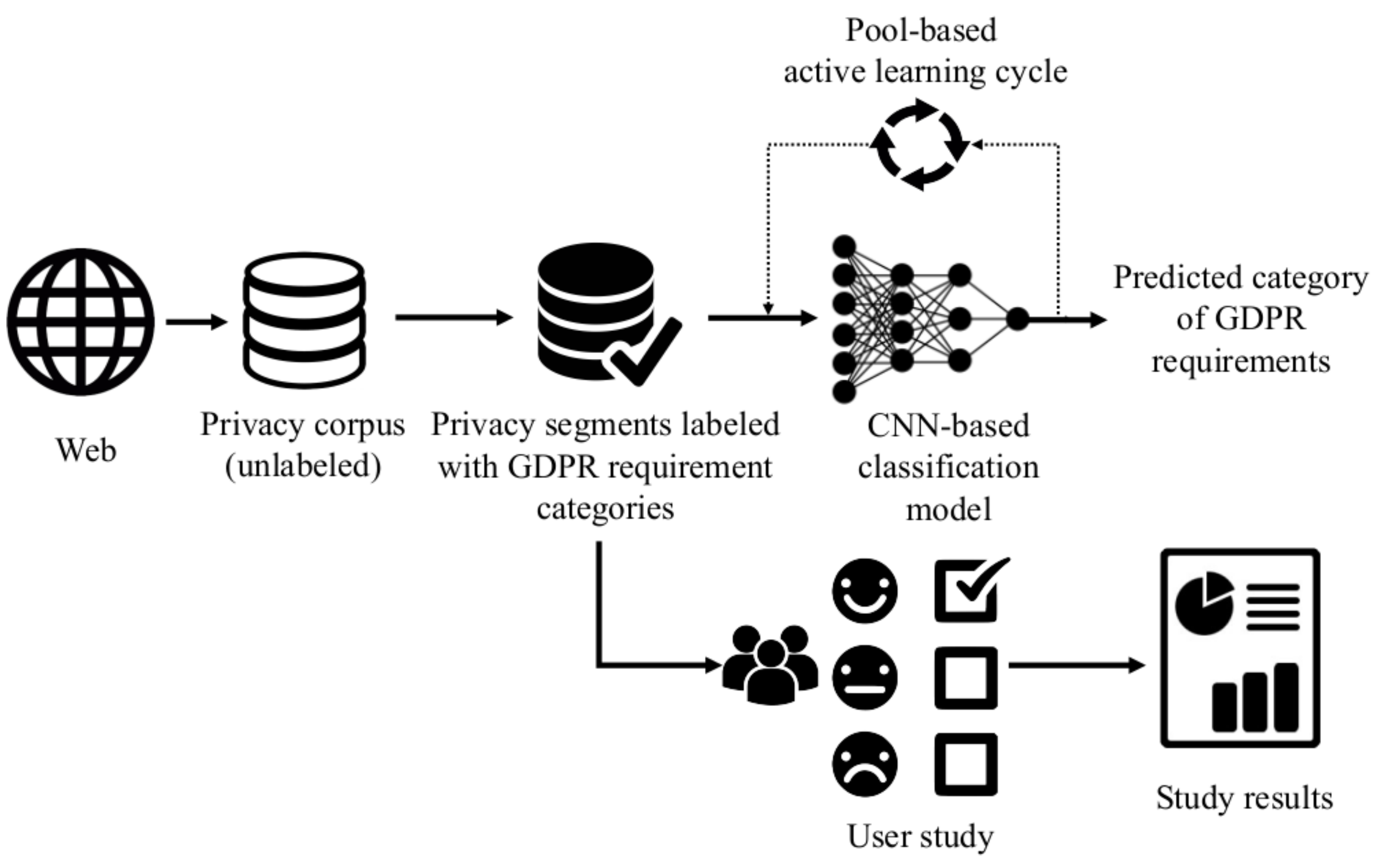}
\caption{Overview of the CNN-based classification model and user perceived effectiveness measurement for the privacy policies relevant to GDPR requirements.}
\label{fig:overview}
\end{figure}

Figure ~\ref{fig:overview} gives a high-level overview of our approach. We first create a privacy policy dataset with the GDPR requirements and then train a CNN-based classification model using our dataset. We further improve the performance using active learning. Additionally, we conduct a comprehensive user study to measure the effectiveness of privacy policies from users' perspective. 

\subsection{Classification Model Overview}
To train our supervised model for classifying GDPR requirements, we create a labeled privacy policy dataset containing 9510 policy segments from 1080 privacy policies. To build the dataset, we collect plain privacy policies from 9761 most visited websites by using UK's IP addresses as UK is one of the top English speaking countries in EU. From 9761 plain privacy policies, we randomly select 1080 policies to be annotated by  trained annotators. 
Upon consolidating the annotation from four annotators, we finally annotate 9510 privacy policy segments with 18 GDPR requirements, which we use to train our CNN-based classification model.

We train a CNN-based model to classify the segments in a privacy policy into 18 classes, each of which represents a disclosure requirement of GDPR. Before training the model, we represent the input texts with sparse low-dimensional vector (embeddings), which gives the CNN-classification model more generalization power. To make the embeddings specific to privacy policy, we initially train an unsupervised word embedding model using FastText~\cite{fasttext} with our entire privacy corpus that contains 9761 privacy policies. Then, for training the classification model, we split our labeled dataset containing 1080 privacy policy documents into 864 (80\%) as training data and 216 (20\%) as test data. Our trained CNN-based model initially achieves an accuracy of 80.5\% with an average F1-score as 0.79. We find the primary reason for misclassifications is the overlapping features between classes. For example, \textit{``You have the right to object, in relation to specific processing of your personal data''-Selectminds.com} represents an instance from \textit{Right to Object} class, whereas \textit{``You have the right to request that we delete your personal data''-Spotify.com} represents an instance from \textit{Right to Erase} class. Both privacy policy segments describe users' right and personal data, although they represent two different classes. To improve the accuracy, the model needs to learn to extract the distinguishable features between the classes that may have overlapping features. One way to resolve this problem is to increase the amount of training data. However, since obtaining more training data is expensive and time-consuming, we use the \textit{active learning} technique to further improve the performance of the model while using less amount of training data.


The primary hypothesis in active learning is that if a learning algorithm can choose the data it can learn most effectively from, instead of learning from a large pool of randomly sampled data, it can perform better than traditional learning methods with substantially less amount of training data. We use iterative pool-based active learning method since obtaining unlabeled privacy policy data is free in our case. In particular, in each iteration, we randomly select a set of unlabeled privacy policy documents from the corpus and feed the unlabeled privacy policy segments into the trained classification model. Then, we use margin sampling to select the instances (queries) that the model predicted with the least confidence. We manually label those ambiguous instances by following the annotation approach similar to what we did to build the original dataset. Finally, we re-train the classification model with the newly labeled training data, which helps the model discriminate between the classes with overlapping features more effectively. We repeat this iterative active learning approach until the overall performance of the model no longer improves. Overall, we achieve an accuracy of 89.2\% with an average F1-score of 0.88, while adding only 10.5\% additional training data.

With the compliance classification model, we run a measurement study for the compliance of the top 10k websites' privacy policies.


\section{Background} \label{sec:background}
In this section, we provide definitions of terminologies used in GDPR requirements and describe the key requirements that are compulsory for privacy policies.

\subsection{GDPR terminologies}
\noindent\textbf{Data Subject.} GDPR (Art. 4) defines data subject as any person whose personal data is collected, stored or processed by data controllers. Personal data can refer to anything from a person's name, address or social media information. Thereby, anyone can become data subject when they book a flight, apply for a job or use a credit card for the purchase. For the convenience of the readers, we use \textit{'user'} to refer to data subject throughout the paper.\\
\textbf{Personal data.} GDPR (Art. 4) defines personal data as any information that relates to an identified or identifiable individual (data subject). Information such as name, address, ethnicity, gender, and web-cookies can be considered as personal data.\\ 
\textbf{Data Controller.} GDPR (Art. 4) defines data controller as any organization, person, or body that controls the personal data and determines the purposes and means of processing data, and is responsible for it, alone or jointly. In short, the data controller controls the procedures and purpose of data usage. For the convenience of the readers, we use \textit{company} to refer data controller throughout the paper.\\
\textbf{Data Processor.} GDPR (Art. 4) defines a data processor as any third party that processes personal data on behalf of a data controller. Throughout the paper, we use \textit{third party} to refer data processor.
    
\subsection{GDPR: Privacy Disclosure Requirements}
We have gleaned through the official GDPR legislation~\cite{gdpr_legislation}, with the help from two legal experts, to find out the information that must be provided in the privacy policy. We find that when personal data are collected from a human subject or other sources, data controllers have to provide 18 categories of information in the privacy policy. More details about these requirements can be found in GDPR regulation (Chapter 3, Art. 12--23)~\cite{gdpr}. In the following, we describe the 18 categories of GDPR requirements with examples of privacy policy segments that companies provide to comply with the requirements.

\textbf{(1) Data Categories}: No matter whether data is collected from users or any other sources, privacy policies should provide the information about the categories of personal data collected, stored or processed by any organization or third parties.

\textbf{(2) Processing Purpose}: Privacy policy should include the purposes of processing user's personal data as well as lawful basis to process those data. Before any organization begin to process user's data, they must determine their lawful basis, which in most cases require that the data processing is \textit{necessary} for specific purposes.

\textbf{(3) Data Recipients}: An organization should disclose the information about the recipients to whom users' personal data have been or will be shared, including the recipients from third countries or international organizations. For example, \textit{Msn.com} provides the information regarding recipients of user's personal data as following - \textit{``We may share your personal information with third parties, such as advertisers, sponsors, and other promotional and business partners.''}

\textbf{(4) Source of Data}: Privacy policy should mention information about which source personal data originate from if they are not obtained directly from users. For example, \textit{TheGuardian.com} discloses the source of some personal information as following - \textit{``We may obtain information about you from partners so that we can make our online advertising more relevant.''}

\textbf{(5) Provision Requirement}: Privacy policy should include whether providing personal data is required, or if the user is obligated to provide the personal data and consequences for not doing so. For example, \textit{Gettyimages.com} discloses the existence of this right in their privacy policy as following - \textit{``You may always choose not to provide personal data, but if you so choose, certain products and services may not be available to you.''}

\textbf{(6) Data Safeguards}: If the controller intends to transfer data to a third country or an international organization with the absence of an adequacy decision, reference of appropriate safeguards of personal data should be provided in the privacy policy. Additionally, privacy policy also needs to provide the means to obtain a copy of the available safeguards to protect users' personal data. 

\textbf{(7) Profiling}: GDPR has provisions on profiling or other automated-decision making. Automated decision making is defined as making a decision solely by automated means without any human involvement, and profiling, which can be a part of automated-decision making, is defined as automated processing of personal data to evaluate certain things about an individual. According to (Art. 22) of GDPR, data controllers can carry out such decision-making if they have a lawful basis and explicit consent of users for the relevant processing of personal data. The privacy policy should include the existence of any automated-decision making, including profiling and what information is used for such decision-making.

\textbf{(8) Storage Period}: Privacy policy should disclose the period for which personal data will be stored, or if not possible, the criteria used to determine that period. GDPR mandates that personal data shall not be kept longer than is necessary for the purposes for which it is processed. Period of personal data stored by a data controller should be limited to a strict minimum and a time limit for the deletion of the data should be determined and disclosed in the privacy policy by the data controller. 

\textbf{(9) Adequacy Decision}: When controllers intend to transfer personal data to a third country or international organization, the existence of absence of an adequacy decision by the European Commission must be disclosed in the privacy policy. Adequacy decision made by the EU Commission is based on whether a country outside of EU offers an adequate level of data protection. For example, Oracle discloses this requirement as following - \textit{``If personal information is transferred to an oracle recipient in a country that does not provide an adequate level of protection for personal information, oracle will take measures designed to adequately protect information about you, such as ensuring that such transfers are subject to the terms of the EU model clauses.''}

\textbf{(10) Controller's Contact}: Data controllers shall provide their identity and contact details along with the contact details of any controller's representative, if applicable. 
    
\textbf{(11) DPO Contact}: Every organization that collects and process the personal information of EU citizens have to appoint a Data Protection Officer (DPO) and contact information of the DPO has to be provided in the privacy policy. 
    
\textbf{(12) Withdraw Consent}: Privacy policy should include the existence of users' right to withdraw consent at any time. For example, \textit{Nextdoor.com} discloses the existence of this right as following - \textit{``If you wish, you can access the content of the (electronic) consent as well as revoke the consent with effect for the future at any time.''}
    
\textbf{(13) Lodge Complaint}: Privacy policy should include the existence of users' right to lodge a complaint with a supervisory authority. For example, \textit{Gofundme.com} discloses the existence of this right in their privacy policy as following - \textit{``You also have the right to lodge a complaint with the local data protection authority if you believe that we have not complied with applicable data protection laws.''}

\textbf{(14) Right to Access}: Data controller should disclose the existence of users' right to request the controller to access or rectify the personal data. According to GDPR (Art. 15), users have right to obtain information about whether their personal data is being collected, used or stored; the categories of the data; with whom the data have been or will be disclosed; whether the data has been or will be transferred to a third country or organization; and how long the data will be stored or processed. For example, legacy.com discloses that \textit{``you may access the information we hold about you anytime via your profile/account.''}
    
\textbf{(15) Right to Erase}: Data controller must disclose the existence of users' right to erase personal data. As mentioned in GDPR (Art. 17), an organization shall have an obligation to erase personal data when data are no longer necessary for the purposes they were collected or stored. In addition to that, if users withdraw their consent any time, corresponding personal data must be erased. 
    
\textbf{(16) Right to Restrict}: Privacy policy should include the existence of user's right to restrict or suppress the processing of their personal data. For example, \textit{Politico.com} discloses this right as following - \textit{"You may also have rights to restrict our processing of your personal data"}
    
\textbf{(17) Right to Object}: Privacy policy should include the existence of user's right to object the processing of their personal data. For example, users have the absolute right to stop their data being used in direct marketing.
    
\textbf{(18) Right to Data Portability}: Privacy policy should include the existence of users' right to receive the personal data in a structured, commonly used and machine-readable format, and right to transfer those data to another data controller. 

\section{Methodology} \label{sec:methodology}
In this section, we first describe our privacy policy corpus and our labeled dataset. Then, we explain the details of our classification model and the active learning process.

\subsection{Privacy Policy Extraction}
To make sure companies have enough time to change their data practices, we collect the privacy policies three months after the effective date, in October 2018, from the top 10,000 websites listed in Quantcast Top UK websites ~\cite{websites_list}. Since we are not located in EU countries, we use UK-based VPNs to capture the privacy policies for EU. To collect the privacy policies, we first use Yandex Search API~\cite{yandex_api} to search for the URL of the privacy policy for each website. During the search, we append the keywords ``privacy policy" with the domain name of the website. From the top 5 search results, we select the URL containing keywords such as privacy, and policy. During this search operation, we limit our search results to English language only. After fetching the URLs, we scrape the HTML pages to extract the privacy policies in plain text, while removing unimportant information such as images and navigation links. However, we exclude URLs that are broken or do not link to the privacy policy page of the corresponding website. For example, our filtered search results for the Wikipedia.com leads to an article~\cite{wiki-privacy-policy} on \textit{Privacy Policy} from Wikipedia instead of Wikipedia's own privacy policy. We manually filter out such cases from the corpus. Finally, we successfully extract 9,761 privacy policies.
\begin{table}[htbp]
\centering
  \small
  \begin{tabular}{l | c}
    
    Privacy Policy Documents & 1080\\
    Total Words & 283,374\\
    Total Classes & 18\\
    \noalign{\smallskip}
    \hline
    \hline
    \noalign{\smallskip}
    Annotated Segments & 9510\\
    Additional Segments for Active Learning & 1001\\
    Total Annotated Segments & 10511\\
    \noalign{\smallskip}
    \hline
    \hline
    \noalign{\smallskip}
    Annotators Per Document & 4\\
    Total Annotators & 6\\
\end{tabular}
\caption{Statistics on the labeled dataset.}
\label{tab:dataset-statistics}
\end{table}

\subsection{Annotation Process}
With the help of two legal experts, we first identify 18 categories of information (shown in Table \ref{tab:category-statistics}) that are required to be disclosed in the privacy policy to fully comply with GDPR. These categories include information about how companies should handle user's personal data and what are users' rights regarding their personal data. 
Note that GDPR requires the companies to present all the 18 requirements even the companies don't have certain data practice (e.g., companies should still clearly indicate they don't share user data rather than not saying anything). To create a privacy policy dataset labeled with GDPR requirements, we randomly select 1,080 privacy policies from our privacy policy corpus. Our legal experts train six human annotators to annotate the privacy policies with GDPR requirements. Each privacy policy document is annotated by four annotators. Each annotator works independently during the annotation process. We design a detailed \textit{annotation schema} to advise our annotators how to label the instances of data correctly and update the schema based on the feedback from the annotators after each annotation phase. We develop a privacy policy annotation tool in Java to help our expert annotators label privacy policies with 18 GDPR disclosure requirements. The tool allows annotators to load a privacy policy and read the policy segment by segment. We remove all HTML tags and non-ASCII characters to improve the readability. At each step, annotators read a segment from the privacy policy and mark it with any of the 18 categories if the segment represents any GDPR requirement. Privacy policy segments that do not represent any particular GDPR requirement are marked as \textit{other} category, which we exclude in our model. 
We finally receive 11,271 annotated privacy policy segments each of which is annotated by four annotators.

To consolidate the labels, we set a 0.75 agreement threshold. For each privacy policy segment, if at least three of the four skilled annotators agree to a label, we consider it as the true label for the corresponding privacy policy segment. Out of 11,271 labeled segments, 8,826 (78.3\%) belong to this threshold. For the privacy policy segments having agreement threshold 0.5 (13.4\% segments), annotators discuss together to find if an agreement over the true label can be reached. We accept 684 segments (6\%) for which annotators can reach an agreement and reject 829 segments (7.3\%) for which they cannot. We discard the segments that have less than 0.5 agreement threshold: no skilled annotators can reach an agreement on the true labels. As a result, we accept the true labels of in total 9,510 privacy policy segments from 1,080 privacy policies. Table~\ref{tab:dataset-statistics} shows the descriptive statistics for the labeled dataset after consolidation. Table~\ref{tab:category-statistics} shows the category-wise statistics for the labeled dataset, such as the number of labeled privacy policy segments, the mean, and median number of words across the segments in each category.
\begin{table}[htbp]
\centering
  \small
  \begin{tabular}{|l | c | c | c | c |}
    \hline
    \multirow{2}{*}{Requirement Categories} &\multirow{2}{*}{GDPR ref.} & \multirow{2}{*}{Freq.} & \multicolumn{2}{c|}{Words}\\
                                                        \cline{4-5}
                                                        & & & mean & med. \\
    \hline
    1. Data Categories & 14(1.d) & 1443 & 24 & 20\\
    2. Processing Purpose & 13(3) & 1926 & 26 & 22\\
    3. Data Recipients & 13(1.e) & 875 & 26 & 22\\
    4. Source of Data & 14(2.f) & 595 & 26 & 20\\
    5. Provision Requirement & 14(5.b) & 542 & 27 & 25\\
    6. Data Safeguards & 14(1.f) & 331 & 24 & 23\\
    7. Profiling & 14(2.g) & 381 & 27 & 16\\
    8. Storage Period & 13(2.a) & 486 & 29 & 27\\
    9. Adequacy Decision & 13(1.f) & 202 & 41 & 37\\
    10. Controller's Contact & 13(1.a) & 308 & 16 & 14\\
    11. DPO Contact &13(1.b) & 530 & 26 & 25\\
    12. Withdraw Consent & 13(2.c) & 510 & 27 & 25\\
    13. Lodge Complaint &13(2.d) & 345 & 31 & 27\\
    14. Right to Access & 14(2.c)& 388 & 17 & 15\\
    15. Right to Erase &14(2.c) & 197 & 19 & 13\\
    16. Right to Restrict & 14(2.c) & 507 & 17 & 12\\
    17. Right to Object & 14(2.c)& 847 & 33 & 27\\
    18. Right to Portability &14(2.c)& 559 & 29 & 28\\
    \hline
\end{tabular}
\caption{18 categories of GDPR privacy disclosure requirements and their corresponding statistics in our dataset. Mean and median were calculated across the population of words in the segments for each categories.}
\label{tab:category-statistics}
\end{table}

\subsection{Privacy Policy Classifier}
Our classification technique consists of two main parts: (1) an unsupervised training to build \textit{Word Embedding} vectors for privacy policies and (2) a CNN-based supervised training as a classifier for GDPR disclosure requirements. We use Word Embedding and CNN-based classification model due to their great success in recent works of text classification~\cite{DBLP:conf/nips/ZhangZL15, DBLP:conf/naacl/Johnson015,DBLP:conf/naacl/PetersNIGCLZ18}. Figure~\ref{fig:model} illustrates the major components of our model.

\subsubsection{Privacy Policy Specific Word Embeddings}
Classical text classification models represent texts in a dataset using metrics such as words and their frequencies. However, such text representation techniques (e.g., Bag of Words) do not consider the position of the words in a sentence and are not capable of capturing the context and semantic meaning of a sentence. For example, \textit{personal information} and \textit{private data} are often used in the same context in privacy policies. Bag of Words considers them as two different phrases, which fails classification tasks in domains such as privacy policy where semantic context is critical. In addition to this, Bag of Words encodes every word in the dataset as a one-hot-encoded vector with the size of the entire vocabulary, which may result in the curse of dimensionality. To solve this problem, we choose Word Embeddings to represent the words in our dataset. A word embedding is a sparse low-dimensional vector representation of a text, which is learned in an unsupervised manner. Words that appear in the same context in the corpus are represented by similar vectors. This gives the neural network model more generalization power since it allows the model to recognize words that are not in the training set, as long as they are in the large corpus used for training the word embeddings model.

To capture the privacy-specific semantic context, we train a custom word embeddings using our privacy policy corpus containing 9,761 privacy policies. For training word embeddings for this corpus, we use \textit{fastText}~\cite{fasttext} skip-gram model. To capture the semantic context, the skip-gram model predicts nearby words given a source word. However, instead of learning vectors for words directly like well-known word embedding model Word2Vec~\cite{DBLP:conf/nips/MikolovSCCD13} or GloVe~\cite{pennington2014glove}, \textit{fastText} represents each word as an n-gram of characters in addition to the word itself. For example, fastText's representation for the word \textit{privacy}, for n=3, is $<pr, pri, riv, iva, vac, acy, cy>$, where the angular brackets represent the beginning and end of the word. This representation allows fastText to assign vectors to the words that were not even seen during the training, which is an advantage for the domain-specific text classification task.

\subsubsection{CNN-based Model Architecture}
\begin{figure}[htbp]
\centering
\includegraphics[width=\linewidth]{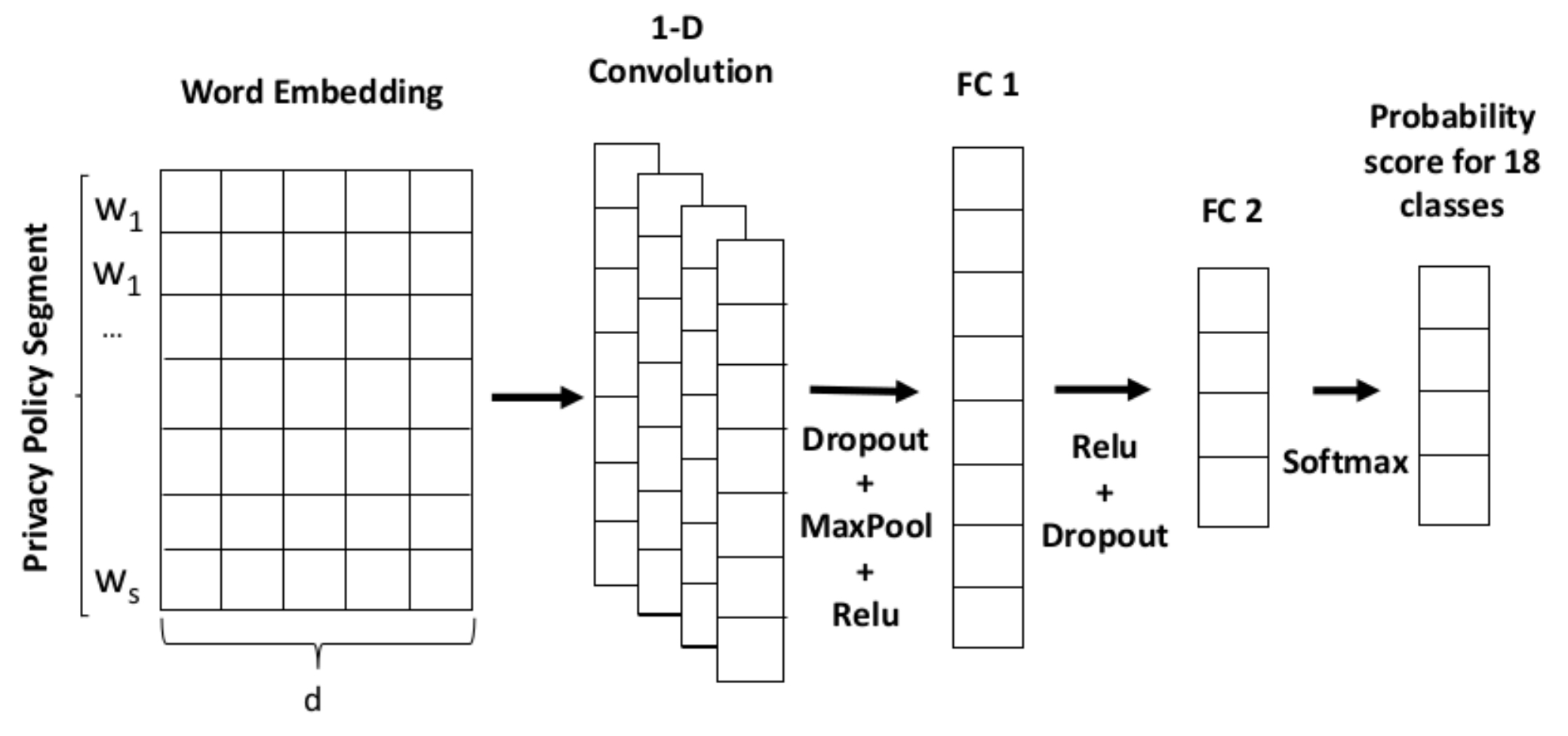}
\caption{Architecture of our Convolutional Neural Network (CNN)-based classification model.}
\label{fig:model}
\end{figure}

We design a CNN-based multi-class classifier to predict the probability of the classes given an input sentence taken from privacy policies. As shown in Figure~\ref{fig:model}, an embedding layer is followed by a one dimensional CNN layer in our classifier model. The CNN layer allows us to use the pre-trained word embeddings, which provides the capabilities to capture semantic meaning from the input sentence. In addition, a CNN layer with filter size $k$ can recognize the features (i.e., set of words) that represent a certain class, regardless of the position of the features in input texts. This layer applies a rectified linear unit (ReLU) as the activation function. Since we use a pre-trained privacy policy specific word embedding, we do not train the embedding layer in our model so that the embedding weights do not get updated while the classifier is learning. To prevent our model from over-fitting, we apply a dropout layer with $0.1$ probability for regularization. A max-pooling layer is applied to extract the most important features from the input.
Output vector from the CNN layer is applied to a fully connected layer, which is again followed by a dropout layer with a probability of $0.5$. This fully connected layer also applies ReLU as the activation function and is followed by another fully connected layer containing the number of units the same as the total number of classes. Finally, we use \textit{softmax} function to determine the probability for each class.
In our problem scenario, collecting labeled data is time-consuming and expensive, whereas collecting unlabeled data (privacy policies) is free. Thereby, to improve the performance of our trained classification model, we leverage pool-based active learning to achieve higher accuracy with less amount of training data.

\subsection{Pool-based Active Learning}
Active learning has been successfully used to improve the accuracy for the machine learning tasks~\cite{DBLP:conf/rep4nlp/ShenYLKA17,DBLP:conf/iclr/SenerS18,DBLP:conf/aaai/ZhangLW17,DBLP:conf/interspeech/DrugmanPK16} where unlabeled data can be easily obtained and labeled data is difficult to collect. The intuition behind using active learning is that the learning model can perform better even with substantially less amount of training data if the learning model can choose the data it needs instead of learning from a large set of randomly sampled data (passive learning). Since privacy segments generally disclose information about users' privacy and rights, it is common for segments from different categories to have overlapping features, which results in misclassification by the model. To overcome this, we use active learning to let our model learn from the segments that the model initially predicted with least confidence between two classes. To select such segments (queries), we first sample instances from our large pool of unlabeled privacy corpus. We classify the sampled instances using our trained classifier. Finally, based on the results from the classifier, we use \textit{margin sampling} to decide whether to label an instance or not.

\begin{figure}[htbp]
\centering
\includegraphics[width=\linewidth]{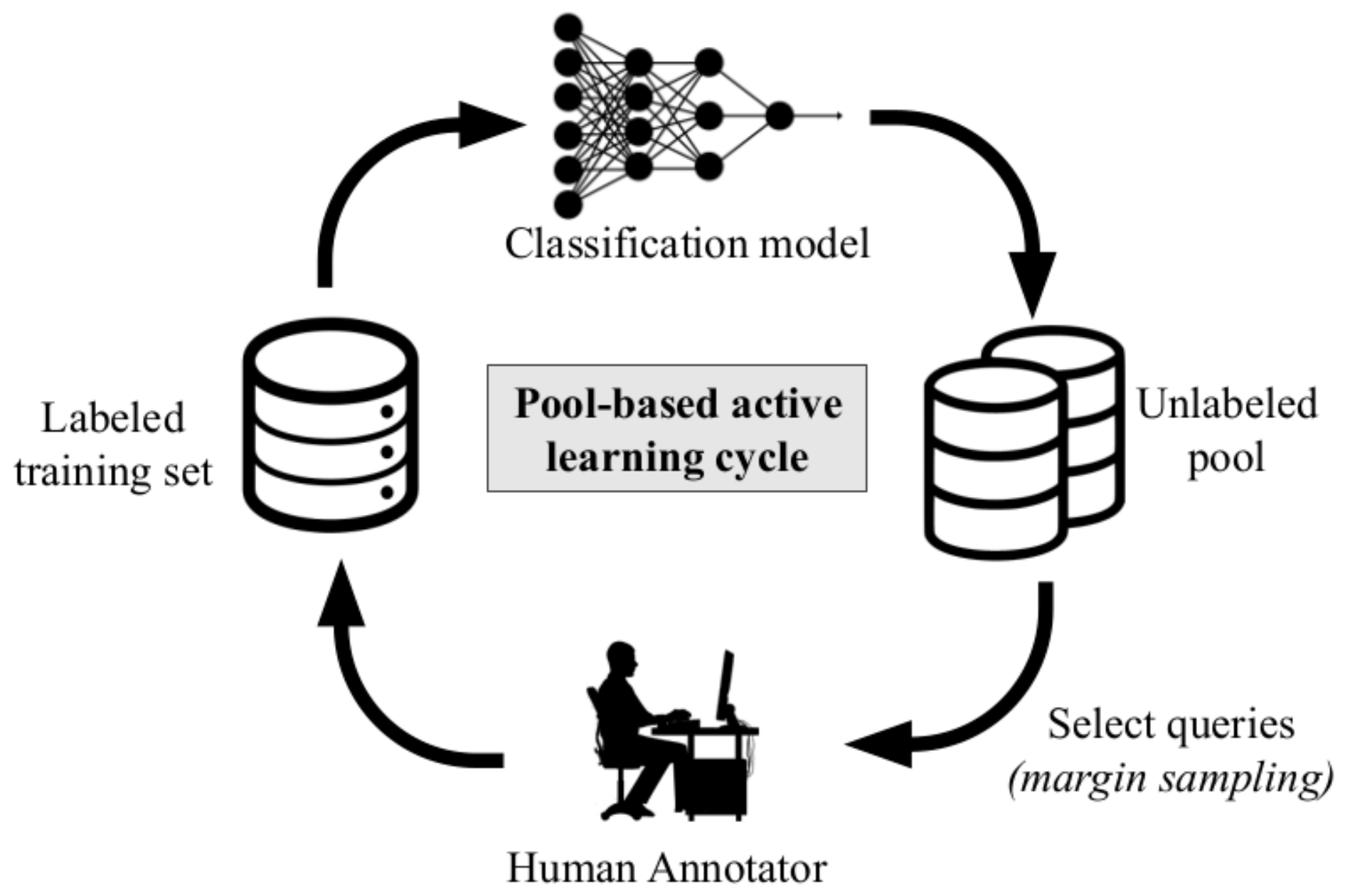}
\caption{Active learning framework to improve classification accuracy.}
\label{fig:active_learning}
\end{figure}

Figure~\ref{fig:active_learning} illustrates the key components of our active learning framework. Since a large pool of unlabeled segments can be obtained at once, we use pool-based selection sampling to sample unlabeled segments from our corpus. We sub-sample a small set of unlabeled privacy policies and pass the segments as input to our classification model. Based on the output probabilities from the model, our active learning framework decides whether to query a segment for labeling or discard it. To this end, we discard any instance that receives a maximum probability score of $0.5$ or less because segments with 0.5 or less probability score are less likely to be relevant to any of the GDPR disclosure requirements. For other instances with more than 0.5 probability scores, we use margin sampling to decide whether to label it or not. Margin sampling considers the difference of prediction score between the first and second most probable classes, based on the probability score of the classification model. We select a query instance that has a minimum difference, which means the model is less confident between the two classes.

\[ x_m = argmin ( P(\hat{y_1}|x) - P(\hat{y_2}|x) )\]
Here, $\hat{y_1}$ and $\hat{y_2}$ represent the probabilities of the first and second most probable classes for segment $x$ predicted by the model. Thus, $x_m$ represents the instances that the classification model predicted with the least confidence. Intuitively, instances with a large margin of prediction scores are easy since the model has little doubt in differentiating between the two most likely classes. On the other hand, instances with small margins are ambiguous. Therefore, knowing the true labels would help the model to discriminate those two classes more efficiently.

\section{Experiments and Results} \label{sec:results}
In this section, we present details of our model development and performance evaluation. We then present the measurement results of using our model to analyze how well the companies are following the GDPR requirements.

\subsection{Model Development}
\noindent\textbf{Data Pre-processing.} We remove all punctuation marks, special characters, digits, and whitespaces from the privacy policy documents. We also remove all stop words since these words are not useful for the learning model and may cause additional memory overhead.

\noindent\textbf{Vector Representation.} We train a fastText word embedding model to encode each word in our privacy policy corpus into a 300-dimensional vector. During the training, we use a minimum length of n-gram as three and maximum length as six, which means the fastText model uses all the substrings contained in a word between three to six characters. This allows the embedding model to encode any new word that is unseen during the training. Overall, we train five epochs with the learning rate of 0.05.

\noindent\textbf{Initial Training and Testing.} We use 864 privacy policies (80\% of the dataset) to train our model. We divide each privacy policy into segments and encode the segments to build a CNN layer, in which we use 400 filters with kernel size of four. We use zero-padding with stride one for the 1-D convolution operation in this layer. The CNN layer uses ReLU activation function and is followed by a dropout layer with the rate of 0.1 and a max-pooling layer. Output vectors from the CNN layer are fed into a fully connected (FC) layer with 256 units. This layer is followed by another FC layer with 18 units. We apply softmax function on the output vector of this layer to determine the probability score for each of the 18 classes in our dataset. After training 50 epochs with the learning rate of 0.001, we test our initial trained model on 216 (20\% of the dataset) privacy policies and achieve an accuracy of 80.5\% with an F1-score of 0.79. 

\begin{figure}[htbp]
\centering
\includegraphics[scale=0.4 ]{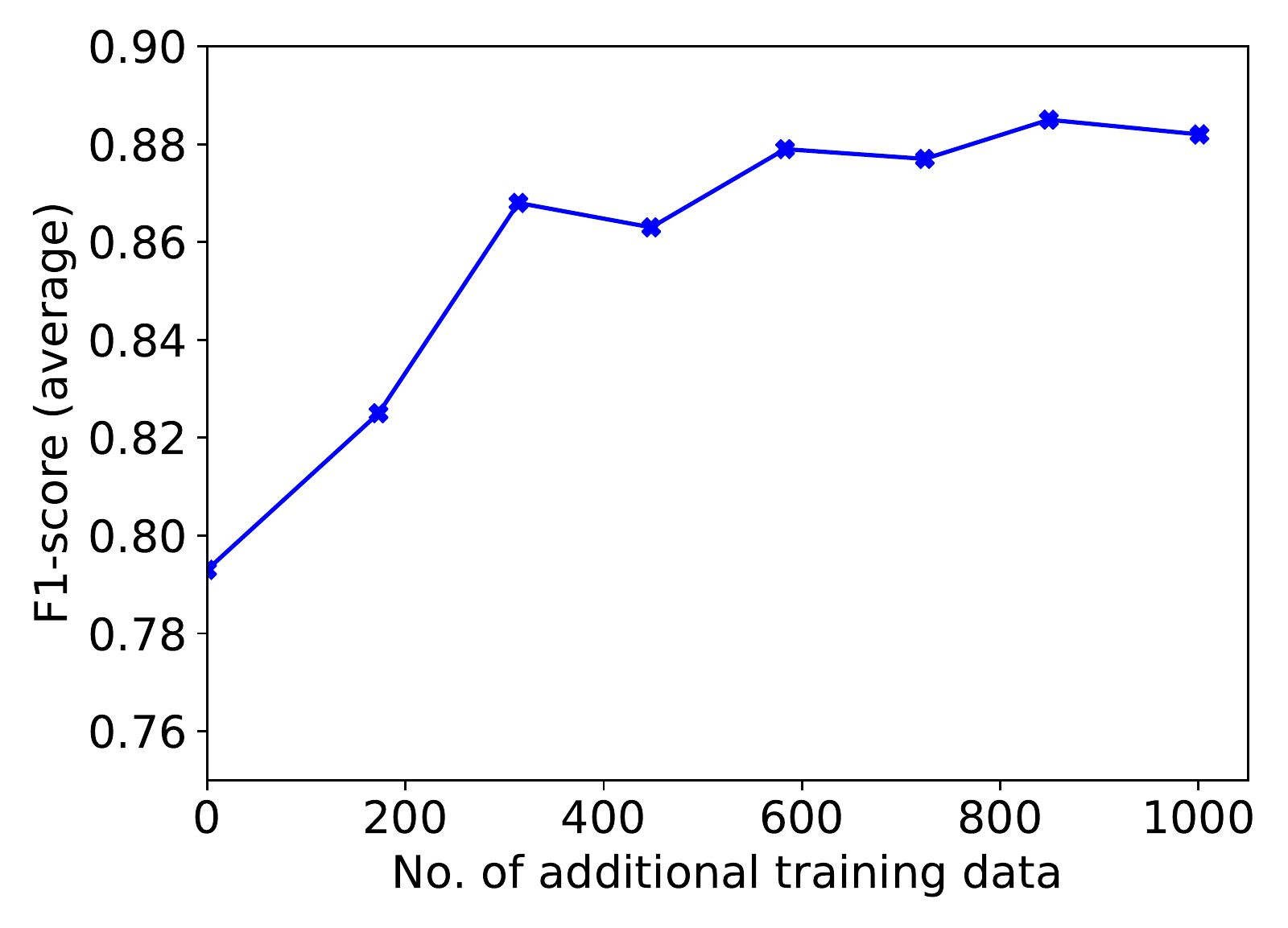}
\caption{Improvement of model's performance during active learning iterations.}
\label{fig_active_learning_f1score}
\end{figure}

\noindent\textbf{Performance improvement.} We manually investigate some misclassified cases and find that the primary reason for a privacy segment being misclassified is because of the existence of overlapping features between the segments. We further use pool-based active learning to make our model more effective in discriminating between the segments from two classes having overlapping features. In each iteration of active learning framework, we randomly sample 100 privacy policies from our large pool of unlabeled privacy policies. Segments from these privacy policies are then fed into our trained classification model. Based on the predicted probability score of the model, we use margin sampling to decide whether to label a segment or not. In each iteration, we select 250 segments that the model predicted with least confidence, particularly with minimum margin between the predicted score of two classes. However, we discard any segments with probability score of 0.5 or less, since such segments are more likely to be irrelevant to GDPR requirements. For annotating the new segments, we follow the similar annotation process and consolidation strategy as discussed in Section~\ref{sec:methodology}. After annotating the new segments, we feed them as training data to re-train the classification model. Overall, we conduct seven iterations until when feeding new training data no longer improves the average performance of our model. Figure~\ref{fig_active_learning_f1score} illustrates the improvement of performance (average F1-score) during each iteration.

\subsection{Performance Evaluation} After conducting seven iterations of active learning, we evaluate our final model using 216 privacy policies (20\% of the labeled dataset) as test set, which are never seen by the classification model during any phase of the training. We achieve an overall accuracy of 89.2\% with average F1-score as 0.88, while adding in total 1001 additional training data, which is only 10.5\% of the initial training data. Table ~\ref{tab:classification-results} presents the precision, recall, and F1-score (macro-average per label) of the evaluation on the test set. As evident in the table, our CNN-based classifier can predict the GDPR requirements from a given privacy policy segment with high accuracy. On average, we achieve 90\% precision, 86\% recall, and 0.88 F1-score. These metrics are higher than the other automated privacy policy analysis tool presented in ~\cite{opp-115, polisis, DBLP:conf/codaspy/TesfayHNKS18}. 
The primary reason for the misclassification in our final model is that the privacy segments might be too vague or might cover multiple requirements with a very brief description.

\begin{table}[htbp]
\centering
  \small
  \begin{tabular}{|l|c|c|c|c|}
    \hline
    Classes & Prec. & Recall & F1 & Support\\
    \hline
    1. Data Categories &      0.87 &     0.75    & 0.81  &   289\\
    2. Processing Purpose &      0.86 &     0.80    & 0.83  &   373\\
    3. Data Recipients &      0.77 &     0.88    & 0.82  &   191\\
    4. Source of Data &      0.87 &     0.77    &  0.82 &   119\\
    5. Provision Requirement &      0.99 &     0.92    &  0.95 &   112\\
    6. Data Safeguards &      0.84 &     0.83    & 0.83  &   70\\
    7. Profiling &      1.00 &     0.88    &  0.94 &   77\\
    8. Storage Period &      1.00 &     0.99    &  0.99 &   83\\
    9. Adequacy Decision &      0.89 &     1.00    & 0.94  &   32\\
    10. Controller's Contact&      0.84 &     0.80    & 0.82  &   66\\
    11. DPO Contact      &      0.97 &     0.98    & 0.98  &   114\\
    12. Withdraw Consent &      0.95 &     0.96    &  0.96 &   85\\
    13. Lodge Complaint &      0.94 &     0.94    &  0.94 &   78\\
    14. Right to Access &      0.82 &     0.77    &  0.80 &   63\\
    15. Right to Erase &      0.90 &     0.76    &  0.82 &   37\\
    16. Right to Restrict &      0.90 &     0.78    &  0.83 &   116\\
    17. Right to Object &      0.95 &     0.83    &  0.89 &   167\\
    18. Right to Portability &      0.86 &     0.92    &  0.89 &   116\\
\hline
    \textit{Average} &      \textit{0.90} &     \textit{0.86}    &  \textit{0.88} &   \\
    
    \hline
\end{tabular}
\caption{Classification results for 18 privacy disclosure requirements of GDPR.}
\label{tab:classification-results}
\end{table}

\subsection{GDPR Compliance Measurement} 

\begin{figure}[!ht]
\centering
\includegraphics[width=\linewidth ]{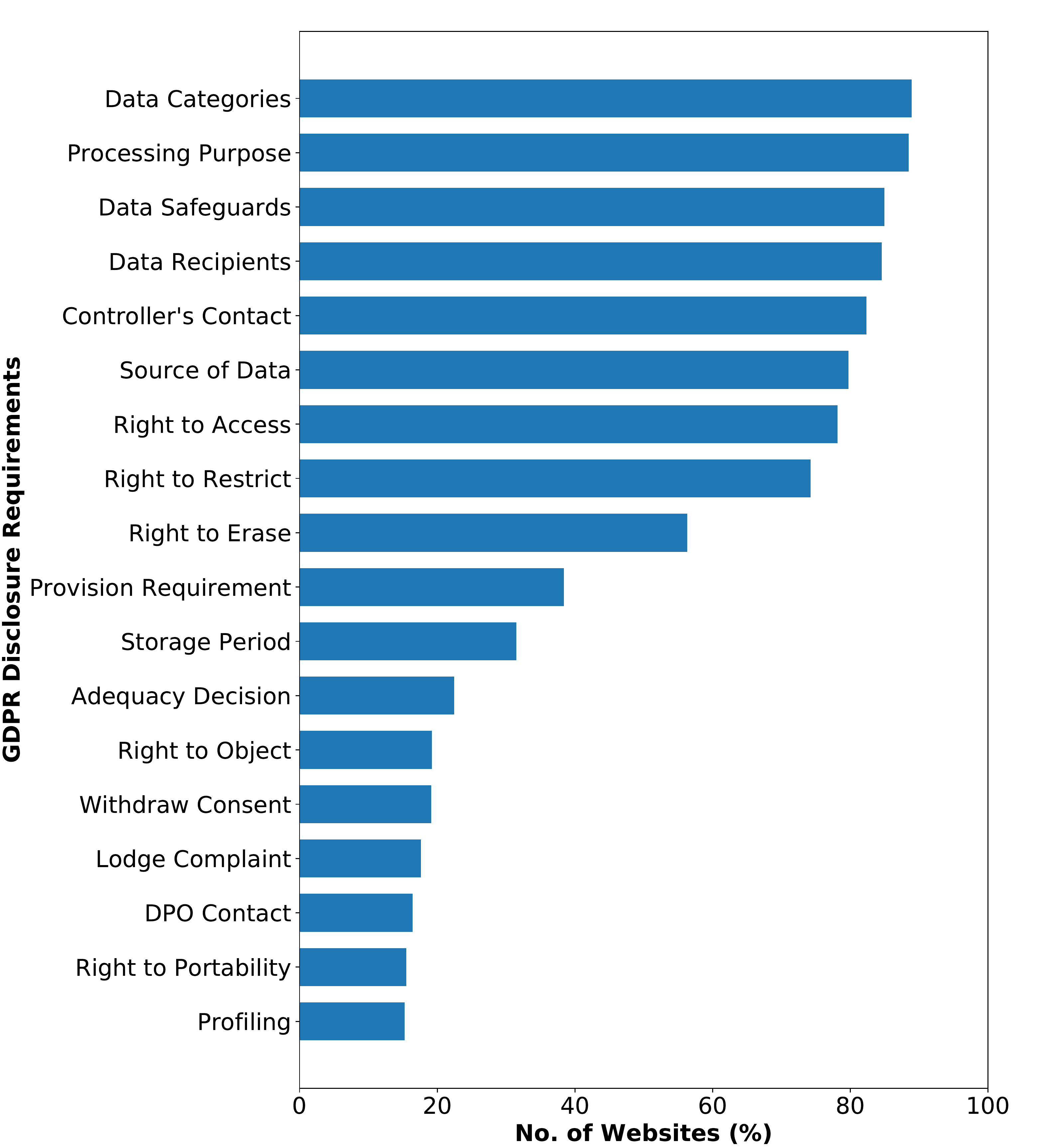}
\caption{Visual representation of the compliance measurement for websites. The measurement result is produced by our CNN-classifier across 9761 websites' privacy policies.}
\label{fig:barchat-compliance-scenario}
\end{figure}

\begin{figure}[!ht]
\centering
\includegraphics[width=\linewidth ]{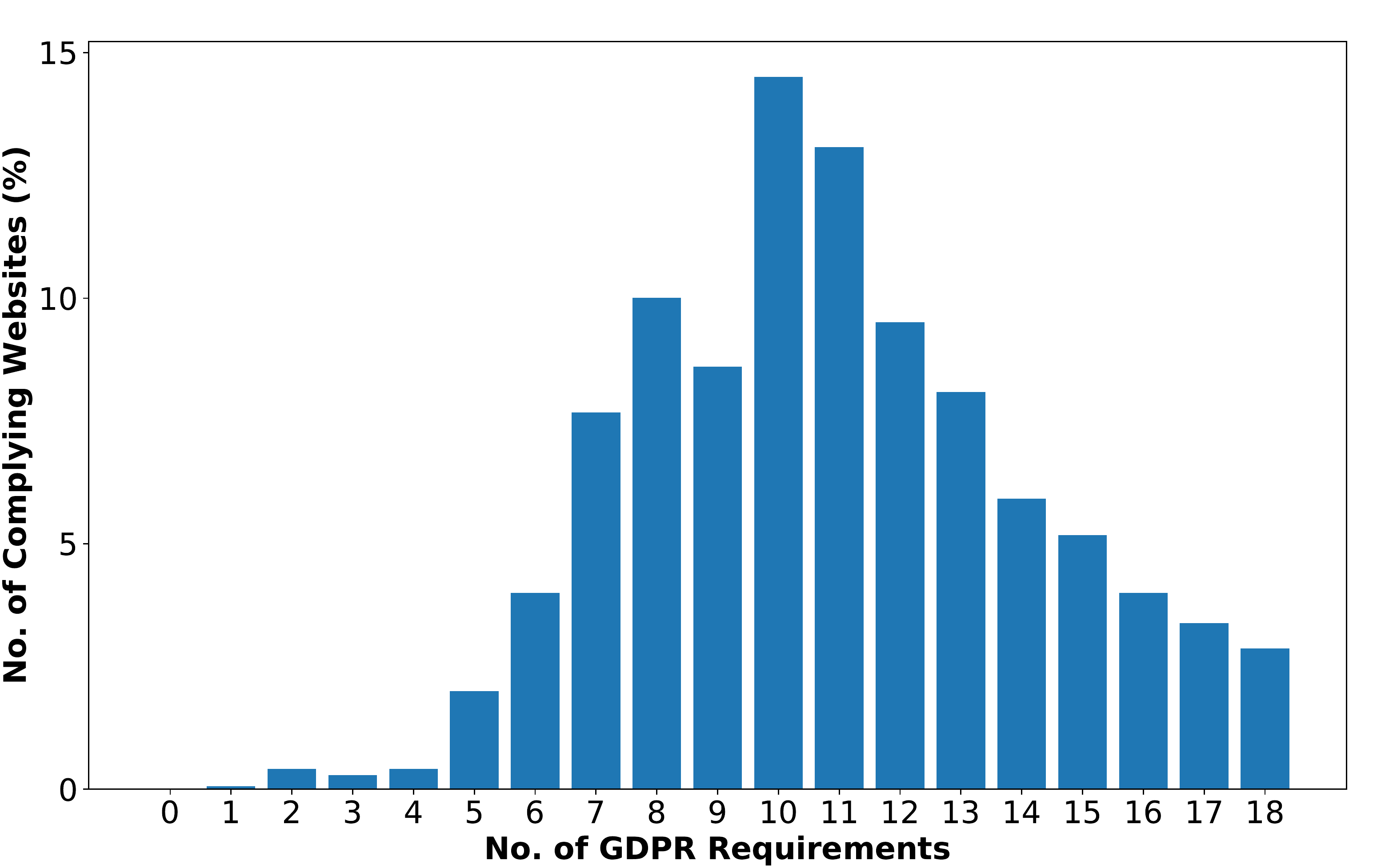}
\caption{Number of websites' policies complying with 0-18 GDPR disclosure requirements.}
\label{fig:barchat-compliance-scenario2}
\end{figure}

We apply our classifier to measure the compliance situations with GDPR requirements of top websites. In particular, to detect how websites comply with GDPR privacy disclosure requirements, our model predicts the GDPR disclosure requirements for each of the privacy policies in the entire corpus (9,761 policies). Figure~\ref{fig:barchat-compliance-scenario} illustrates the number of websites meeting the GDPR privacy disclosure requirements for each of the classes. We find that many companies still do not follow the requirements introduced by GDPR. For example, among the 9,761 websites, only 15.3\% follow the GDPR requirement of \textit{profiling}, which requires that companies have to disclose how users' personal information is used for automated decision making or profiling purpose. Also, requirements such as users' \textit{Right to Portability} and providing the contact information of Data Protection Officer (DPO contact) are covered by 15.5\% and 16.4\% websites, respectively. Other primary requirements, such as users' right to Lodge Complaint, Withdraw Consent, Right to Object, and Adequacy Decision, are covered by 17-20\% websites. Figure~\ref{fig:barchat-compliance-scenario2} shows the number of companies complying with 0-18 GDPR requirements. It appears that only 3\% of websites fully comply with 18 requirements. These findings indicate that many websites still do not follow the requirements of GDPR.

\section{Discussion} \label{sec:discussion}
\noindent\textbf{Limitation.} We only consider English language privacy policies since our team is not familiar with other languages in the EU. However, we are able to perform in-depth analysis to make up for language coverage. In addition, there might be other ways that companies disclose information to users but our work only considers privacy policy since it is a dominating way to communicate privacy disclosures to users~\cite{reidenberg2015disagreeable}. In fact, California Consumer Privacy Act~\cite{ccpa} requires companies to disclose information to users via privacy policies. Moreover, since our CNN-based classifier takes privacy policy document as input, it is out of the scope of this paper to identify whether companies' underlying data-practices are GDPR-compliant. 

\noindent\textbf{Future Work.} 
Further research can be conducted to predict the effectiveness of a privacy policy from human perception. Building such a tool will help the developers understand the usability issues of their privacy policies. In addition, analyzing flow-to-policy consistency in terms of GDPR requirements would be an interesting research problem. 

\section{Related Work} \label{sec:relatedwork}
Privacy policies are one of the most common ways to disclose how companies collect, store or process users' personal data. Researchers have worked on identifying data practices from privacy policies by leveraging both manual and automated approach. Tesfay et al.~\cite{DBLP:conf/codaspy/TesfayHNKS18} built a privacy policy summarization tool based on 11 privacy aspects of GDPR. However, since their tool was built upon pre-defined keywords-based features and trained on only 45 privacy policies, it resulted in lower accuracy and could not provide a reliable large-scale analysis. Degeling et al.~\cite{DBLP:journals/corr/abs-1808-05096} performed a longitudinal study on the privacy policies and cookie consent notices of 6,579 websites representing the 500 most popular websites in the EU countries. They showed that 72.6\% of websites updated their privacy policies close to the date of May 25, 2018 when GDPR came into effect. However, their work primarily focused on cookie consent notices and reflection of GDPR terminologies in privacy policies without considering cross-checking compliance with GDPR requirements. Basin et al.~\cite{basin2018gdpr} proposed a theoretical methodology to decompose GDPR compliance for auditing with the idea of identifying a business process and a purpose. Unlike their work, we contribute an automated tool to assess GDPR compliance in privacy policies using machine learning technique.

Researchers have also performed privacy policy analysis using natural language processing and machine learning. Wilson et al.~\cite{opp-115} introduced a corpus of 115 privacy policy documents annotated with fine-grained data practices (OPP-115). Harkous et al.~\cite{polisis} leveraged OPP-115 to build an automated QA model (Polisis) for regular data practices in privacy policies. Linden et al.~\cite{linden2020privacy} designed seven queries adapted from Polisis to represent limited number of GDPR requirements and performed compliance analysis using filtering and scoring approach. Chang et al.~\cite{chang2019automated} proposed an automated privacy policy extraction system to predict and extract app's privacy policies based on users’ concerns under different contexts. However, these papers focused on extracting fine-grained data practices from privacy policies and did not consider disclosure requirements of the privacy laws. Hence, they cannot be easily extended for assessing the compliance of privacy policies with GDPR, which introduces more rules than regular data practices along with unprecedented number of privacy rights regarding data collection and processing.
\section{Conclusion} \label{sec:conclusion}

In this paper, we provide insights into the compliance and effectiveness of privacy policies regarding the disclosure requirements of GDPR. We first identify comprehensive categories of GDPR requirements for privacy policies and create a privacy policy dataset labeled with 18 GDPR requirements. We then build a CNN-based automated tool to classify segments in a privacy policy into GDPR requirements with an accuracy of 89.2\%. We find that only 3\% of companies fully comply with GDPR in their privacy policies.

\bibliographystyle{plain}
\bibliography{references}
%




\end{document}